\documentclass[aps,prb,showpacs,twocolumn,superscriptaddress]{revtex4}
\usepackage{amsmath}
\usepackage{amssymb}
\usepackage{graphicx}
\begin{document}

\title{Non-magnetic B-site Impurities Induce Ferromagnetic Tendencies in CE Manganites}
\author{Xiao Chen}
\affiliation{Laboratory of Solid State Microstructures, Nanjing University, Nanjing 210093, China}
\author{Shuai Dong}
\affiliation{Laboratory of Solid State Microstructures, Nanjing University, Nanjing 210093, China}
\affiliation{Department of Physics and Astronomy, University of Tennessee, Knoxville, Tennessee 37996, USA}
\affiliation{Materials Science and Technology Division, Oak Ridge National Laboratory, Oak Ridge, Tennessee 32831, USA}
\author{Kefeng Wang}
\affiliation{Laboratory of Solid State Microstructures, Nanjing University, Nanjing 210093, China}
\author{J.-M. Liu}
\affiliation{Laboratory of Solid State Microstructures, Nanjing University, Nanjing 210093, China}
\affiliation{International Center for Materials Physics, Chinese Academy of Sciences, Shenyang 110016, China}
\author{Elbio Dagotto}
\affiliation{Department of Physics and Astronomy, University of Tennessee, Knoxville, Tennessee 37996, USA}
\affiliation{Materials Science and Technology Division, Oak Ridge National Laboratory, Oak Ridge, Tennessee 32831, USA}
\date{\today}

\begin{abstract}
Using a two-orbital model and Monte Carlo simulations, we investigate the effect of nonmagnetic B-site substitution on half-doped CE-type manganites. The lattice defects induced by this substitution destabilize the CE phase, which transforms into (1) the ferromagnetic (FM) metallic competing state, or (2) a regime with short-range FM clusters, or (3) a spin-glass state, depending on couplings and on the valence of the B-site substitution. While a C-type antiferromagnetic state is usually associated with an average $e_{\rm g}$ charge density less than 0.5, the nonmagnetic B-site substitution that lowers the $e_{\rm g}$ charge density is still found to enhance the FM tendency in our simulations. The present calculations are in qualitative agreement with experiments and provide a rationalization for the complex role of nonmagnetic B-site substitution in modulating the phase transitions in manganites.
\end{abstract}
\maketitle

\section{Introduction}
Rare-earth manganites of the form $R_{1-x}A_{x}$MnO$_3$ (where $R$ ($A$) is a rare-earth (alkaline-earth) element) are typical representatives of complex oxides with multi-orbital interactions and a strong competition between spin, charge, orbital, and phononic degrees of freedom.\cite{Tokura:Bok,Dagotto:Prp,Salamon:Rmp} In these materials the main competition is between the delocalization tendency of the $e_{\rm g}$ electrons and the localization effects caused by the antiferromagnetic (AFM) coupling between the Mn $t_{\rm 2g}$ spins as well as the Jahn-Teller effects.\cite{Tokura:Bok,Dagotto:Prp,Salamon:Rmp, Dagotto:Sci} The delicate balance between these competing tendencies produces a very rich phase diagram. Intrinsic or external perturbations that naively may seem ``weak'', such as small variations in the carrier density, pressure, magnetic fields, and quenched disorder, can nevertheless induce nonlinear effects, including phase transitions due to the close proximity in energy of the competing states. This high sensitivity to perturbations is clear in manganites with composition $x=0.5$, where several competing ground states with quite different properties, such as ferromagnetic as well as CE and A-type AFM states, have been identified.\cite{Yunoki:Prl,Hotta:Prl,Kajimoto:Prb,Kagan:Jetp} Among them, the CE state is known to appear  in several narrow-bandwidth manganites. This complex state is formed by zigzag FM chains with AFM inter-chain coupling, and its stabilization is usually accompanied by a checkerboard pattern of charge and orbital order, which further stabilizes the CE spin structure. \cite{Yunoki:Prl,Hotta:Prl,Kajimoto:Prb}

Metal-insulator transitions (MIT) are familiar phenomena in condensed matter physics. For a normal metal, the addition of quenched disorder leads to the trapping of mobile carriers and transforms a metal into an insulator. However, for manganites the {\it reversed} process usually occurs, in situations where quenched disorder originates from chemical substitution or intrinsic defects.\cite{Tokura:Rpp} Usually, two sources of quenched-disorder effects in manganites are considered. One is primarily caused by the A-site disorder (alloy randomness),\cite{Wang:Prb,Wang:Apl} while the other is induced by B(Mn)-site substitution. \cite{Sakai:Prb,Barnabe:Apl,Raveau:Jssc,Martin:Jmmm,Yaicle:Prb,Yaicle:Jssc,Markovich:Prb,Nair:Prl,Banerjee:Prb,Hardy:Prb,Nucara:Prb} Theoretically, the A-site disorder effects have been extensively studied in the past few years. Bond disorder and on-site potential disorder were jointly or separately introduced to model various A-site disorder sources. \cite{Aliaga:Prb,Motome:Prl,Alvarez:Prb,Yu:Prb,Sen:Prb,Sen:Prb06,Sen:Prl,Burgy:Prl,Salafranca:Prb,Kumar:Prl,Kumar:Prl06,Kumar:Prb,Kumar:Prl08}
The quantitative calculations indicate that the FM metal phase is not much affected by the A-site disorder, while the CE/CO insulator easily collapses into a glassy-like state. 

Conceptually, and  differently from the A-site disorder, the B-site substitution induces disorder locally directly into the Mn-O network and it may modulate quantities such as $n$, the $e_{\rm g}$ electron-density. Therefore, the B-site substitution can have stronger impact on the physical properties of the material than the A-site disorder. Experiments show that a few percent B-site substitution, such as chemical substitution of Mn by Cr/Al/Ga/Ru in the half-doped CE state of manganites, will favor a phase separated (PS) state with FM tendencies, although these ions are different in their electronic structure and have different magnetic coupling with the Mn ions. \cite{Barnabe:Apl,Raveau:Jssc,Martin:Jmmm,Yaicle:Prb,Yaicle:Jssc,Markovich:Prb,Nair:Prl,Banerjee:Prb,Hardy:Prb} For the nonmagnetic substitution, for instance, $2.5\%$ Al substitution in Pr$_{0.5}$Ca$_{0.5}$MnO$_3$ is sufficient to convert the charge-ordered (CO) CE insulator into a state with FM metallic characteristics.\cite{Banerjee:Prb} These experimental results are very surprising since (1) the substituting ions Al/Ga are nonmagnetic, and (2) the Al/Ga doping leads to the reduction of the $e_{\rm g}$ electron density, both of which are disadvantageous for the FM tendency.\cite{Nucara:Prb} In fact, such a reduction of the $e_{\rm g}$ electron density is expected to favor an AFM insulating state.\cite{Nucara:Prb}

Very recently, the disorder effects by B-site substitution were also investigated within the context of the two-orbital double-exchange (DE) model.\cite{Pradhan:Prl,Pradhan:El} The observed collapse of the CE/CO phase into a FM phase was explained in terms of a density-driven phase-separation. The main idea is that the impurity, having a $+4$ valence, transfers extra electrons to the remaining Mn sites.\cite{Pradhan:El} Thus, in this context it is straightforward to understand the FM tendency since $e_{\rm g}$ densities larger than half, $n_{\rm r}$$>$$0.5$, usually are associated with FM phases in manganites. However, this density-driven phase-separation idea can not explain the experimental fact that {\it trivalent} substitutions, such as Cr/Al/Ga, are expected to reduce the $e_{\rm g}$ electron density on the remaining Mn-site ($n_{\rm r}$$<$$0.5$) rather than increasing this density. This reduction of the $e_{\rm g}$ density is quite nontrivial since the electronic density is one of the most important factors to determine the ground state, especially around the half-doping region. In fact, the previous theoretical investigation predicted that Al/Ga would $not$ lead to a FM tendency for half-doped manganites based on the previously described idea of a density-driven phase-separation.\cite{Pradhan:El} Therefore, it is necessary to re-investigate the B-site-substitution disorder effects in half-doped manganites to understand these puzzling experimental results.

In the following, we will study the effects of B-site substitution on the stability of the CE/CO state. Since even Al/Ga substitutions can induce the FM tendency, it is reasonable to consider first a non-magnetic impurity for simplicity. The main results found in this paper are that the B-site disorder can induce two main effects: (1) The lattice defects by B-site substitution break the CE-zigzag chains and frustrate the charge ordering which destabilizes the CE phase and induce the FM tendency (the FM state is close in energy to the CE phase in half-doped manganites); (2) The reduction from half-doping of the $e_{\rm g}$ electron density suppresses both the long range FM and CE tendencies, since it prefers the C-AFM state. These two competing effects can trigger a phase transition from the CE/CO phase either into a short-range FM cluster state with relatively strong FM tendencies, or into a spin glass state. Due to these competing effects, an optimal B-site substitution level for the stabilization of the FM tendency is found to exist.

\section{Model}
In our investigations, we consider a two-orbital model defined on a two-dimensional $L\times L$
square lattice ($L=8$, and using
periodic boundary conditions) with a Hamiltonian given by,
\begin{eqnarray}
\nonumber H&=&-\sum_{<ij>,\alpha,\beta,\sigma}t^{{v}}_{\alpha\beta}d_{i\alpha\sigma}^{\dagger}d_{j\beta\sigma}+J_{\rm AF}\sum_{<ij>}\textbf{S}_{i}\cdot \textbf{S}_{j}\\
&&\nonumber-J_{\rm H}\sum_{i}\textbf{s}_{i}\cdot\textbf{S}_{j}+\lambda\sum_{i}(Q_{1i}\rho_{i}+Q_{2i}\tau_{xi}+Q_{3i}\tau_{zi})\\
&&+\frac{1}{2}\sum_{i}(2Q_{1i}^2+Q_{2i}^2+Q_{3i}^2)-\mu\sum_{i} n_i,
\end{eqnarray}
where the first term is the two-orbital DE interaction, $d_{i\alpha\sigma}^{\dagger}$ ($d_{i\alpha\sigma}$) is the creation (annihilation) operator for an
$e_{\rm g}$ electron with spin $\sigma$ in the orbital $\alpha$ ($d_{x^{2}-y^{2}}$ or $d_{3z^{2}-r^{2}}$) at site $i$. The hopping amplitudes between nearest-neighbor (NN) sites $<ij>$ 
are given by $t_{aa}^{x}=-\sqrt{3}t_{ab}^{x}=-\sqrt{3}t_{ba}^{x}=3t_{bb}^{x}=1$ for $v=x$, and $t_{aa}^{y}=\sqrt{3}t_{ab}^{y}=\sqrt{3}t_{ba}^{y}=3t_{bb}^{y}=1$ for $v=y$. The second term is the AFM super-exchange (SE) interaction between the NN $t_{\rm 2g}$ spins $\textbf{S}$. In the third term, the Hund coupling $J_{\rm H}$ ($>0$) links the $e_{\rm g}$ electrons with the $t_{\rm 2g}$ spins $\textbf{S}$ (assumed classical and
normalized as $|\textbf{S}|=1$). For simplicity, we consider here the Hund coupling in the widely-used
limit of $J_{\rm H}\rightarrow\infty$. The fourth term is the electron-phonon coupling, where $\lambda$ is the dimensionless coupling constant, $Q$ are the
phononic modes ($Q_1$ is for the breathing mode, $Q_2$ and $Q_3$ are for the Jahn-Teller modes), and $\tau$ is the orbital pseudospin operator. The fifth term in the Hamiltonian is the elastic energy of the phonons.\cite{Hotta:Prl03} For simplicity, and as in many other previous investigations, the phonons here will actually be considered just as classical lattice distortions. $\mu$ in the last term is the chemical potential to tune the $e_{\rm g}$ electron's density.

To introduce a non-magnetic impurity (without $3d$ electrons) as the B-site substitution, we will assume that the impurity has no contribution to the electron conductivity and, thus, consider it as a lattice defect. Thus, the DE, SE, and Jahn-Teller couplings around the impurity can be ignored, retaining only the elastic energy of local phonons. These localized defects distinguish the B-site substitution models \cite{Pradhan:Prl, Pradhan:El} from the A-site disorder models \cite{Aliaga:Prb,Motome:Prl,Alvarez:Prb,Yu:Prb,Sen:Prb,Sen:Prb06,Sen:Prl,Burgy:Prl,Kumar:Prl,Kumar:Prl06,Kumar:Prb,Kumar:Prl08} in which the disorder effects were applied to all sites. Therefore, the topological structure of the lattice defects, which is absent in A-site disorder cases, is especially important in patterning the electron configuration of the original CE/CO state.

Our model Hamiltonian Eq.~(1) is studied via a combination of exact diagonalization and Monte Carlo (MC) techniques: classical $t_{\rm 2g}$ spins and phonons evolve following the MC procedure; and at each MC step, the fermionic sector of the Hamiltonian is numerically exactly diagonalized. The first $10^{4}$ MC steps are used for thermal equilibrium and another $10^{3}$ MC steps are used for measurements. More details about this widely-used two-orbital Hamiltonian and the MC algorithm can be found in Ref.~\onlinecite{Dagotto:Prp}. In the present calculation, first we considered averages over several defect configurations. However, we observed that for a dilute distribution of defects, namely with defects not in close proximity to each other,  the results of the calculations are almost the same for different configurations. Therefore, here only two defect configurations were used for each parameter point in most simulations, except for the cases of the phase diagram and density-of-states (DOS) for which only one configuration was used. In addition, since defects should have the same probability to occupy the two types of sites in the CE phase, the bridge sites (B1) and corner sites (B2),\cite{Dong:Prb} the same number of defects on these B1 and B2 sites are arranged in our $8\times8$ lattice. The $8\times8$ lattice is enough to describe the prominent phenomena of the B-site substitution, such as the destabilization of the CE order, the phase separation, and the FM tendency, since the lattice size effects are mild in these phenomena.\cite{Motome:Prl,Sen:Prb06} All the simulations are performed at a low-temperature fixed at $T$=$0.01$, which is low enough to describe ground state properties. To characterize different spin orders, the spin structure factors are calculated by performing Fourier transforms of the real-space correlation functions.\cite{Dong:Prb08}

\section{Results}

\begin{figure}
\vskip -0.6cm
\centerline{\includegraphics[width=10cm]{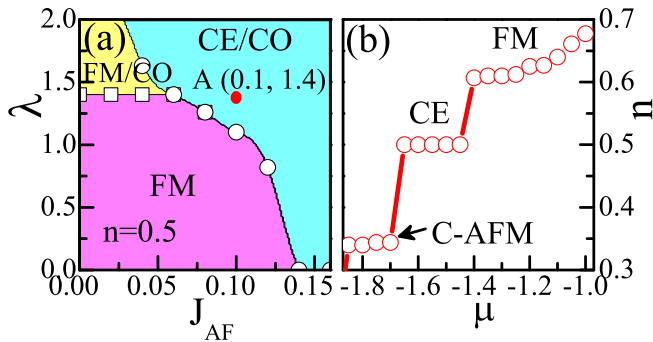}}
\vskip -0.7cm
\caption{(Color online) (a) $J_{\rm AF}$-$\lambda$ phase diagram at $x$=$0.5$ in the absence of B-site substitution (clean limit). Here, point A refers to the particular set of couplings $J_{\rm AF}$=$0.1$ and $\lambda$=$1.4$ that is emphasized in our analysis. (b) The $e_{\rm g}$ charge density vs. chemical potential $\mu$, at $J_{\rm AF}$=$0.1$ and $\lambda$=$1.4$ (point $A$ in Fig.1~(a)).}
\vskip -0.5cm
\end{figure}

\subsection{Clean limit phase diagram}
First, we will briefly review the phase diagram of the model used here when the A-site doping is $x=0.5$, and in the clean limit. The result is shown in Fig.~1(a). This phase diagram in the ($J_{\rm AF}$, $\lambda$) plane can be divided into three main regimes. When both $J_{\rm AF}$ and $\lambda$ are small, the DE interaction dominates and favors the FM metallic state (regime denoted by ``FM''). With increasing $J_{\rm AF}$, i.e. enhanced SE interaction, an appropriate combination of $J_{\rm AF}$ and $\lambda$ leads to the CE/CO insulating state (regime ``CE/CO''). Moreover, there exists a parameter-space region with coexisting FM order and charge order (regime ``FM/CO'') where $J_{\rm AF}$ is small and $\lambda$ is large. This phase diagram was established before and verified experimentally, thus the reader is referred to previous literature for more details.\cite{Tokura:Bok,Dagotto:Prp,Salamon:Rmp}  Here our attention will concentrate on the CE/CO regime near the boundary with the FM metallic phase. Considering point A ($J_{\rm AF}=0.1$ and $\lambda=1.4$) as an example (in the rest of the paper, $J_{\rm AF}$ is fixed to 0.1 unless otherwise stated), Fig.~1(b) presents the ground state density in the vicinity of $n=0.5$, as a function of the chemical potential $\mu$. A clear plateau with $n=0.5$ indicates a fairly stable CE phase. The other two plateaus correspond to two other phases: the FM state at $n>0.5$ and the C-type AFM state at $n<0.5$, indicating the importance of the charge-density variation in driving the phase transition. The transitions between the three phases are abrupt when varying the chemical potential $\mu$, suggesting density-variation-driven first-order phase transitions, at least in the small clusters we have studied in this effort.\cite{Dagotto:Prp,Yunoki:Prl,Hotta:Prl,Kumar:Prl,Pradhan:Prl,Dong:Prb08}

\subsection{Effect of lattice defects}
Now let us investigate the effect of B-site nonmagnetic substitution in manganites of the form $R_{0.5}A_{0.5}$Mn$_{1-y}B_{y}$O$_3$. Such a substitution will lead to the appearance of lattice defects and, simultaneously, a variation of the $e_{\rm g}$ electron density. To clarify their respective roles, here we first address the effect of the lattice defects. For such purpose, the substituting cations are assumed to be $+3.5$ in charge to keep the average $e_{\rm g}$ charge density for the remaining Mn-sites ($n_{\rm r}$) invariant, i.e. $n_{\rm r}=0.5$.

We have observed that for a given appropriate substitution level $y$, the CE/CO state will turn into a state with strong FM tendency. Fig.~2(a) shows several typical spin structure factors $S(\textbf{q})$ evolving with $y$ at the point ($J_{\rm AF}$=$0.1$, $\lambda$=$1.4$). The FM order at $\textbf{q}$=$(0,0)$ emerges at $y>0.031$ and it is enhanced quickly up to $S(\textbf{q})$$\sim$$0.12$. On the other hand, the E-type AFM order at $\textbf{q}$=$(\pi/2,\pi/2)$ and C-type AFM order at $\textbf{q}$=$(0,\pi)$ are rapidly suppressed when $y$ reaches $0.063$, indicating that the CE spin order is destroyed by the lattice defects. This result is qualitatively similar to the result in Fig.~3(d) of Ref.~\onlinecite{Pradhan:Prl}. However, the origin of this CE to FM transition is by lattice defects (to be explained below), instead of the $e_{\rm g}$ density enhancement proposed in Ref.~\onlinecite{Pradhan:Prl}. The CE destruction can be further understood by observing the MC snapshot of a spin configuration at $y$=0.094 substitution. As shown in Fig.~2(b), there is no trace of any CE chains, the CE phase is converted into a state consisting of small FM clusters with various orientations, similar to the results in previous A-site disorder efforts. \cite{Alvarez:Prb}

\begin{figure}
\vskip -0.6cm
\centerline{\includegraphics[width=10.5cm]{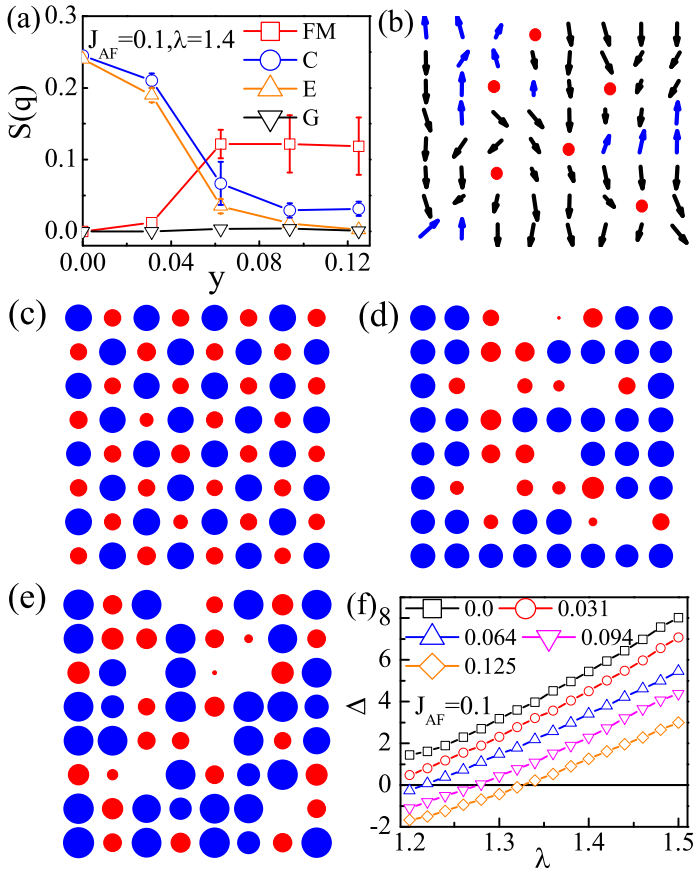}}
\vskip -0.6cm
\caption{(Color online) (a) Spin structure factor $S(\textbf{q})$ for various spin orders as a function of $y$, at $J_{\rm AF}$=$0.1$ and  $\lambda$=$1.4$. (b) Typical MC snapshot of the spin configuration showing the short range FM domains ($y$=0.094, $J_{\rm AF}$=$0.1$, $\lambda$=$1.4$), where the substitution ions are represented by filled circles. This is in the ``FM cluster'' regime (R2) discussed in the text. (c-e) Typical MC snapshot of the charge distribution. Here the circle area is proportional to the local charge density. High- and low-density sites (comparing with $0.5$) are colored by blue and red, respectively. (c) The staggered CO pattern in the clean limit ($J_{\rm AF}$=$0.1$, $\lambda$=$1.3$). (d) CO pattern at  $y$=$0.094$, $J_{\rm AF}$=$0.1$, and $\lambda$=$1.3$. (e) CO pattern at  $y$=$0.094$, $J_{\rm AF}$=$0.1$, and $\lambda$=$1.6$, in the spin-glass regime (R1). The locations of the substituting ions are the same in (b), (d) and (e). (f) Energy difference between the FM and CE ordered states as a function of $\lambda$, for various values of $y$.}
\vskip -0.6cm
\end{figure}

The fundamental reason for the CE/CO destabilization can be understood based on the breaking of the zigzag FM chains and concomitant charge frustration. Considering first the magnetic order, the CE phase consists of zigzag FM chains that are easily cut down by lattice defects, leading to a substantial increase in the kinetic energy. However, the competing FM phase has a two-dimensional (three-dimensional in real case) character, which is much more robust against lattice defects. For a pure system the Jahn-Teller coupling favors the long-range staggered CO pattern.  In real cases, the B-site substitution should be randomly distributed between the B1 sites (with higher charge density) and B2 sites (with lower charge density with the same probabilities. This randomness of B-site substitution can break the original CO state, causing charge frustration. This frustration will spread over the whole lattice and it leads to the collapse of the long-range charge order. Two typical MC snapshot of the charge redistribution are shown in Fig.~2(d) and (e) with intermediate and large $\lambda$ values. For the intermediate $\lambda$ case (here $\lambda$=$1.3$), the charge density distribution is homogeneous except for some regions around the defects, while for the large $\lambda$ case (here $\lambda$=$1.6$), the charge disproportionation is obvious although it occurs without a long-range ordered pattern.

The intuitive idea described in the previous paragraph can be examined by means of a crude calculation: the phonons (classical distortions) are allowed to evolve freely in the MC sequence, while the $t_{\rm 2g}$ spin background is {\it frozen} into either the FM or CE patterns. The energy difference between the FM and CE phases, denoted by $\Delta$ defined as $\Delta$=$E_{\rm FM}-E_{\rm CE}$, is shown in Fig.~2(f). In the clean limit $y$=0, the energy difference $\Delta$ is relatively large, but it is rapidly suppressed upon increasing $y$. An appropriate choice of $y$ and $\lambda$ may allow the energy of the FM phase to be even lower than that of the CE phase ($\Delta$$<$$0$), and in these cases the ground state favors the FM order. This crude calculation illustrates the asymmetric impact of lattice defects on the stability of the two phases, suggesting a possible transition from the CE/CO phase to the FM phase. In fact, previous theoretical studies on the A-site disorder effects have also confirmed the fragility of the half-doped CO phases.\cite{Motome:Prl,Sen:Prb,Sen:Prb06,Burgy:Prl,Kumar:Prb,Kumar:Prl08,Aliaga:Prb,Alvarez:Prb}

\begin{figure}
\vskip -0.6cm
\centerline{\includegraphics[width=10cm]{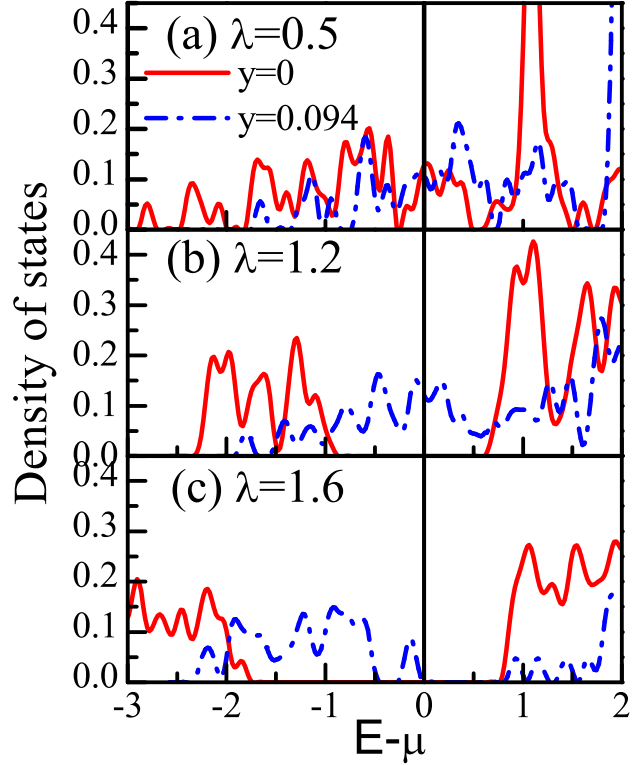}}
\vskip -0.6cm
\caption{(Color online) (a)-(c) Calculated $e_{\rm g}$ electron density-of-states, DOS, for the two values of $y$ indicated, at $\lambda$=$0.5$, $1.2$, and $1.6$. $J_{\rm AF}$ is fixed to 0.1.}
\vskip -0.6cm
\end{figure}

Let us study now the electronic structure and transport properties of the system. The $e_{\rm g}$ electronic DOS provides insight on the effect of lattice  defects. Figs.~3(a)-(c) show
the calculated DOS at several values of $y$ and $\lambda$, 
for a fixed $J_{\rm AF}$=$0.1$. The DOS at small
$\lambda$ ($\lambda$=$0.5$) is not qualitatively modified by the lattice defects, showing the anticipated
robustness of the metallic state. For a large electron-phonon coupling $\lambda$=$1.6$, the DOS shows a large energy gap at the
Fermi level in the clean limit $y$=$0$, corresponding to the long range CE/CO phase. A substitution
of $y$=$0.094$ clearly shrinks this gap, but still there are no states at the Fermi level (although there are some states close to it).
However, for an intermediate coupling $\lambda$=$1.2$, the most exotic
features in the DOS are obtained. In the clean limit, the gap is wide and obvious, but this gap
completely vanishes at $y$=$0.094$, suggesting the stabilization of a finite DOS at the Fermi
level and, if Anderson localization is not considered, 
metallic behavior in the electronic transport. This lattice-defects-induced
insulator to metal transition is similar to that induced by the A-site
disorder.\cite{Motome:Prl,Sen:Prb,Sen:Prb06,Burgy:Prl,Kumar:Prb,Kumar:Prl08,Aliaga:Prb,Alvarez:Prb}

\begin{figure}
\centerline{\includegraphics[width=10cm]{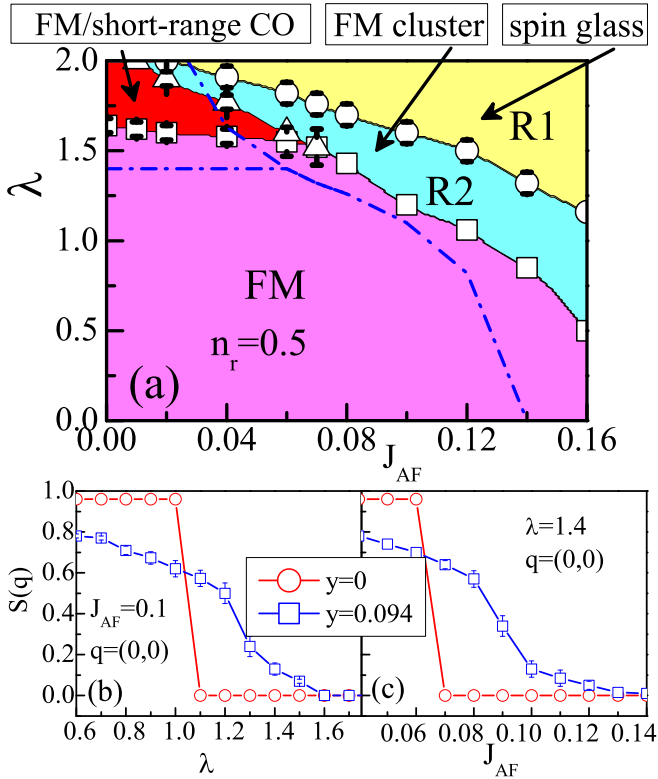}}
\vskip -0.4cm
\caption{(Color online) (a) $J_{\rm AF}$-$\lambda$ phase diagram at $y$=$0.094$ compared with the clean-limit phase diagram Fig.~1 (a) (the
blue lines are the
phase boundaries in Fig.~1(a)). (b) Spin structure factor $S(\textbf{q})$ at $\textbf{q}$=$(0, 0)$ as a function of $\lambda$, for the
 two substitution levels indicated. (c) Spin structure factor $S(\textbf{q})$ at $\textbf{q}=(0, 0)$ as a function of $J_{\rm AF}$, for the
 two substitution levels indicated.}
\vskip -0.4cm
\end{figure}

As a compact summary of the conclusions of this section, the MC
calculated phase diagram at $y$=$0.094$ is shown in Fig.~4(a).
Comparing with the phase diagram in the clean limit (see Fig.~1(a),
and the blue lines in Fig.~4(a)), the FM metallic phase remains fairly stable
and even expands slightly. This increase in the range of stability of the FM metal 
is important to rationalize
the experimental results of Ref.~\onlinecite{Banerjee:Prb}. 
An interesting feature of the phase diagram Fig.~4(a) is that the CE/CO regime fully
vanishes, with the original boundary with the FM regime shrinking
backward slightly. The original clean-limit CE/CO phase collapses into three sub-regimes:
the FM metal and the two regions
denoted in the figure by R1 and  R2. Here, R1 (large $J_{\rm AF}$
and $\lambda$) corresponds to a ``spin glass regime'' with short-range
charge order and no visible FM order parameter $S(0,0)$. This regime
was described in previous publications, such as
Ref.~\onlinecite{Aliaga:Prb}. In the more novel regime R2, the
microstructure consists of short-range FM clusters with
various orientations (Fig.~2(b)). The charge order is suppressed
in this regime (Fig.~2(c)), and this region is here called the ``FM
cluster regime''. With increasing $J_{\rm AF}$ and $\lambda$, these
FM clusters/domains will be separated into even smaller domains, and
eventually into a spin glass state. For this regime, there is
relatively strong FM tendency, and a visible drop in $S(0,0)$
(by varying $\lambda$ or $J_{\rm AF}$ as shown in Figs.~4(b) and
(c)) is observed when the parameters ($\lambda$ and $J_{\rm AF}$) 
cross the phase boundary between FM metallic and FM cluster regimes. 
In addition to their influence on the modification of
the phase diagram, the lattice defects also smear the phase
boundaries, implying inhomogeneous tendencies in the FM cluster
state. As shown in Figs.~4(b) and (c), the FM order parameter
$S(0,0)$ decreases slowly with increasing $\lambda$ or $J_{\rm AF}$,
in contrast to the abrupt drop characteristic of the first-order
FM-CE transition in the clean limit. {\it Summarizing,} 
the B-site lattice defect disorder replaces the clean-limit CE/CO phase by three
different regimes: (i) A simple extension in parameter range of the competing
FM metallic phase. In Figs.~4(b,c), this regime is between the $\lambda$ or $J_{\rm AF}$ 
where the original first-order
jump from FM to CE in the clean limit occurs, to the clearly visible change in the slope
of the $S(0,0)$ curve (cusp) with further increasing couplings. (ii) The next regime
is the FM cluster state (R2), already described. (iii) The following is the 
spin-glass state (R1), also described
before in detail in this section. Thus, we predict that half-doped manganites in the
CE state could be destabilized in three different manners by B-site defects 
depending on how close they are to the
FM metallic state in the clean-limit phase diagram. 
Qualitatively, these results are similar to the effects caused by
A-site disorder, and this similarity is a consequence of 
the previously unveiled fragility of the CE phase.\cite{Alvarez:Prb}

\subsection{Effect of electronic-density variations}

In the previous section, we have only considered the effects of the lattice defects introduced by the
non-magnetic substitutions, which gave rise to results similar to those of previous studies that focused on A-site
disorder. Now let us incorporate the effect of the $e_{\rm g}$ electron-density variation due to the
B-site substitution. For the pure system, $n$$>$$0.5$ usually corresponds to FM order, while $n$$<$$0.5$
corresponds to the C-type AFM order, as shown in Fig.~1(b). After the substitution, this
density-variation-driven phase transition is still relevant.\cite{Pradhan:El,Dong:Prb08} If the
substituting ions have charge higher than $+3.5$ ($n_{\rm s}$$>$$3.5$), corresponding to $n_{\rm r}$$>$$0.5$,  the
spins present a FM tendency.\cite{Pradhan:El,Dong:Prb08} However, if $n_{\rm s}$$<$$3.5$, the situation
becomes much more complex since $n_{\rm r}$$<$$0.5$ usually leads to the C-type AFM order instead of the FM
one. For example, if the impurity is Cr/Al/Ga with charge $+3$ ($n_{\rm s}$=$3$), $n_{\rm r}$ becomes less than
$0.5$ and the effective electronic density will decrease to \textbf{$(0.5-y)/(1-y)$} with increasing $y$. For $y$=$3.1$$\%$,
$6.3$$\%$, $9.4$$\%$, and $12.5$$\%$, the $e_{\rm g}$ electron-density $n_{\rm r}$ in
$R_{0.5}A_{0.5}$Mn$_{1-y}B_{y}$O$_3$ drops down to $0.484$, $0.467$, $0.448$ and $0.429$,
respectively. Therefore, the reduction of electronic density will compete with the FM tendency induced by the
lattice defects discussed in the previous section.

Let us recalculate $S(\textbf{q})$ for different orders as a
function of $y$, at $J_{\rm AF}$=$0.1$ and $\lambda$=$1.35$, as
shown in Fig.~5(a). Remarkably, a significant FM tendency is still
present and the maximum $S(0,0)$ appears at $y$$\sim$$0.094$, which
can be considered as the optimal substitution level for the FM
order. In contrast to the much reduced C-type AFM order in
Fig.~2(a), here $S(0,\pi)$ is partially sustained due to the charge
density reduction, while on the other hand $S(\pi/2,\pi/2)$ vanishes quickly,
suggesting the rapid disappearance of the CE order. A typical MC
snapshot of the spin configuration considering the $e_{\rm g}$
charge-density reduction effect is shown in Fig.~5(c), where both the FM tendency and C-type AFM tendency are observed simultaneously, namely there are pairs of FM spin chains coupled antiferromagnetically 
as in the C state, and also small pockets of ferromagnetism. 
More specifically for the FM
component, the calculated FM fraction in the spin structure factor
in our simulations is about $17.5$$\%$, which is consistent with the
experimental low-$T$ magnetization of
Pr$_{0.5}$Ca$_{0.5}$Mn$_{0.975}$Al$_{0.025}$O$_3$:
$0.7\mu_{B}/f.u.$, corresponding to $\sim$$20$$\%$ of the saturated
magnetization.\cite{Banerjee:Prb} It should be mentioned that the
above predicted optimal value of $y$ is higher than the
experimentally identified one. This disagreement may be ascribed to
the A-site disorder already existing in Pr$_{0.5}$Ca$_{0.5}$MnO$_3$
and other defects which are not considered in the present model.
Although our results do not seem quantitatively accurate in this
respect, we are confident that the qualitative tendencies have been
captured in our calculations.

\begin{figure}
\vskip -0.4cm
\centerline{\includegraphics[width=10cm]{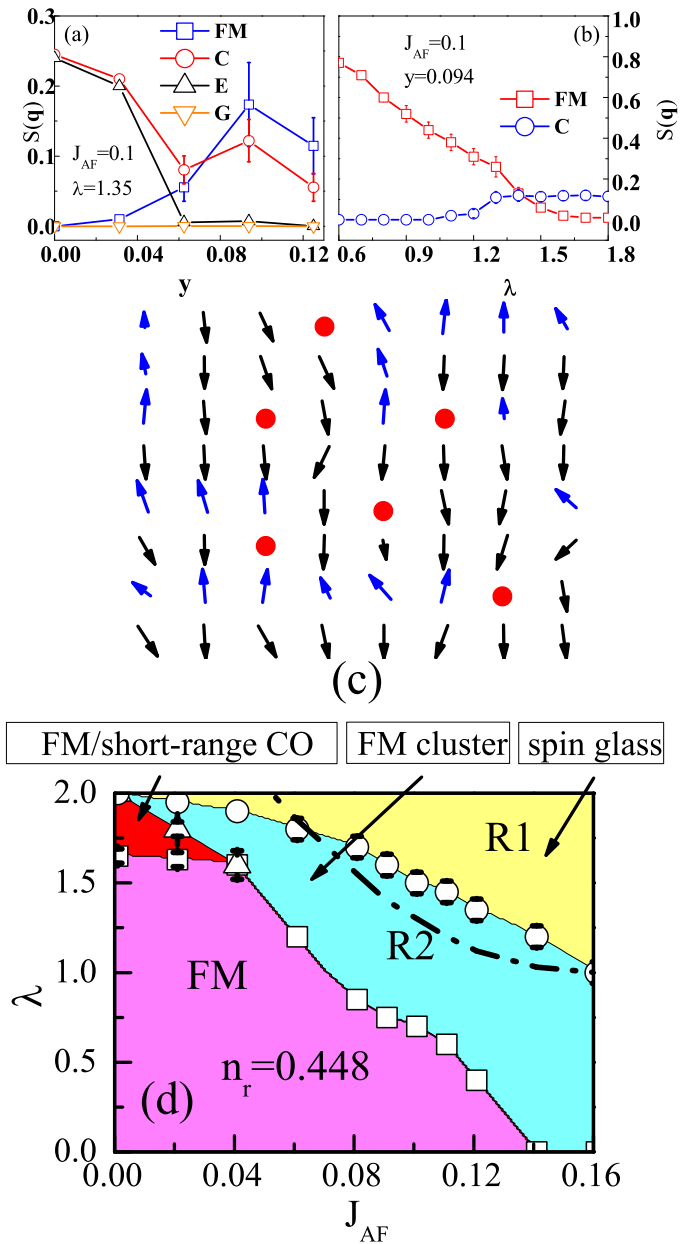}}
\vskip -0.8cm
\caption{(Color online) (a) Spin structure factor $S(\textbf{q})$
for various spin orders as a function of $y$ ($\lambda$=$1.35$)
after taking into account the $e_{\rm g}$ charge-density reduction
effect. (b) Spin structure factor
$S(\textbf{q})$ for two spin orders as a function of $\lambda$. 
(c) Typical MC snapshot of the spin configuration
considering the $e_{\rm g}$ charge-density reduction effect at
$J_{\rm AF}$=$0.1$ and $\lambda$=$1.35$, where the substitution ions
are represented by filled circles. 
(d) $J_{\rm AF}$-$\lambda$ phase diagram at $y$=$0.094$ with
$n_{\rm r}$=$0.448$. The dashed dot-line divides the phase diagram
into two regimes, with the region of stronger $\lambda$ and $J_{\rm AF}$
having a stronger C-type AFM signal.}
\vskip -0.5cm
\end{figure}

As a conclusion of this section, the MC calculated phase diagram at
$y$=$0.094$ with $+3$ impurity cations has also been calculated,
and it is shown in Fig.~5 (d). Comparing with Fig.~4(a), the effect
of the electron density reduction is clear: the FM metallic regime shrinks
while the FM cluster regime is relatively enlarged. For
the regime above the dashed dot line, including portions of R1 and
most of the R2 regime, the C-type AFM component remains robust (see
$S(0,\pi)$ in Fig.~5(b)). In addition, the FM order $S(0,0)$ in
Fig.~5(b) decreases more smoothly with increasing $\lambda$ compared
with the two curves in Fig.~4(b), indicating that the lattice
defects and charge density reduction will both smear the phase
boundary. Thus, the
reduction of $e_{\rm g}$ electron-density suppresses both the
long-range CE and FM spin orders and enhances the importance of the C-AFM order.
Therefore, the total combined effect of lattice defects and electron
density reduction over the clean-limit CE state leads to an inhomogeneous state with coexistence
of short-range FM and C-AFM spin ordering.

\section{Conclusions}

Our Monte Carlo investigations reported here have shown that
the role of the non-magnetic B-site
substitution in manganites is rather complex. There are two main tendencies
that compete: (1) The B-site substitution introduces
 lattice defects that
break the CE-zigzag chains and causes charge frustration; (2) It  also
varies the $e_{\rm g}$ electron-density that leads to a
density-variation-driven phase transition. In principle, both of
these two roles are absent in the A-site disorder case. However, 
the lattice defects appear to induce similar results as the case of
A-site disorder: part of the original CE phase regime in
the clean-limit phase diagram is
taken over by the FM metallic phase, or short-range FM clusters. This can be
understood in the context of the previously discussed ``fragility''
of the CE phase,\cite{Alvarez:Prb} as compared with the robustness of the
FM order. However, the concomitant reduction of $e_{\rm g}$
electron density suppresses {\it both} the FM and CE spin orders 
leaving behind a large
inhomogeneous area that consists of coexisting
short-range FM and C-AFM clusters.
Furthermore, the competition between lattice defects and electron
density reduction gives rise to an optimized substitution level for
the FM tendency, qualitatively similar as found experimentally.

Summarizing, here we have investigated extensively the CE/CO
destabilization experimentally observed in half-doped manganites due to a small
amount of B-site nonmagnetic substitution, by using the two-orbital
double-exchange
model. Our calculations have shown that the CE/CO phase can be
easily destabilized by lattice defects, leading to a variety of interesting possible
states that include the competing FM metal, or FM clusters with or without C-AFM
regions (depending on the valence of the B-site substitution), or a spin glass state.
In particular, the surprising FM tendency observed here driven
by the nonmagnetic substitution into the CE state 
is consistent with
several recent experimental results.

\section{Acknowledgments}
We thank S. Kumar, K. Pradhan, P. Majumdar, and A. Kampf for careful reading and useful comments. This work was supported by the Natural Science Foundation of China (50601013, 10674061), the National Key Projects for Basic Research of China (2006CB921802, 2004CB619004), and the 111 Programme of MOE of China (B07026). S.D. and E.D. were also supported by the NSF grant DMR-0706020 and the Division of Materials Science and Engineering, U.S. DOE, under contract with UT-Battelle, LLC.

\end{document}